\documentclass[reprint,superscriptaddress,amsmath,amssymb,aps,floatfix]{revtex4-2}

\usepackage{graphicx}
\usepackage{dcolumn}
\usepackage{bm}
\usepackage{amsfonts}
\usepackage{amsthm}
\usepackage{gensymb}
\usepackage{url}

\begin{document}

\title{A generative pre-trained transformer with Kerr-soliton attention}

\author{Lindell M. Williams}
\email{lindell.williams@colorado.edu}
\affiliation{Time and Frequency Division, National Institute of Standards and Technology, Boulder, CO 80305, USA}
\affiliation{Department of Physics, University of Colorado, Boulder, CO 80309, USA}
\author{Yan Jin}
\affiliation{Time and Frequency Division, National Institute of Standards and Technology, Boulder, CO 80305, USA}
\affiliation{Department of Physics, University of Colorado, Boulder, CO 80309, USA}
\author{Scott B. Papp}
\affiliation{Time and Frequency Division, National Institute of Standards and Technology, Boulder, CO 80305, USA}
\affiliation{Department of Physics, University of Colorado, Boulder, CO 80309, USA}

\begin{abstract}
Generative pre-trained transformers (GPTs) have enabled capability-rich language models, but their operation incurs substantial costs in computation, memory, and data movement. A major source of this cost is attention, a nonlinear transformation that computes context-dependent weights from token correlations.  
Here, we introduce Kerr-soliton attention, harnessing driven--dissipative nonlinear dynamics in a resonator to realize, execute, and validate deep-learning attention in physical hardware. We train a GPT language model using an analytic representation of Kerr-soliton attention and explore generative inference by streaming model-produced inputs through a time-multiplexed soliton ensemble, whose nonlinear evolution transforms the scores into attention weights. We observe high-fidelity agreement between the experimentally produced weights and those predicted by the model. By mapping memory and nonlinear computation to physical dynamics, Kerr-soliton attention points toward hybrid digital--physical systems that reduce data movement in deep learning.
\end{abstract}

\maketitle

Recently, we have seen rapid advances in machine learning and AI, driven by architectural innovations \cite{vaswani_attention_2017}, expanding datasets \cite{brown_language_2020, gao_pile_2020}, and the continued scaling of model size and compute capacity \cite{brown_language_2020, kaplan_scaling_2020, chowdhery_palm_2023}. Yet this scaling has exposed a bottleneck: for many transformer-based AI systems, a dominant cost is the movement and staging of model parameters and cached states in memory \cite{dao_flashattention_2022, kwon_efficient_2023}. Modern digital accelerators support an extraordinary peak compute exceeding $10^{15}$ operations per second at an energy cost on the order of 1~pJ per operation, while memory bandwidth remains limited to the $\mathrm{TB/s}$ scale. These memory transactions contribute substantially to resource consumption and are a growing point of criticism for long-term sustainability \cite{strubell_energy_2019, patterson_carbon_2021}. Consequently, a range of approaches has been explored to mitigate data movement, including quantization \cite{gholami_survey_2022}, memory-aware algorithm design \cite{dao_flashattention_2022, dao_flashattention-2_2023}, and specialized accelerators \cite{jouppi_-datacenter_2017, chen_eyeriss_2017}. 
However, the reliance on memory traffic remains, motivating alternative approaches that can alleviate memory and compute burdens through different system paradigms.

Among emerging approaches, computing with physical systems offers a path in which operations are realized through the evolution of a dynamical system, rather than through stages of digital computing \cite{momeni_training_2025}.
This perspective is particularly relevant to deep learning, where models are defined by compositions of nonlinear transformations. Photonics provides a compelling resource for machine-learning computations, supporting high-bandwidth signal transport and multiplexed processing through propagation and interference, and has been demonstrated as a viable platform for analog classical computing \cite{mohseni2022ising, inagaki2016}. Linear photonic systems, including specialized, high-capacity laser sources \cite{rizzo_massively_2023, pirmoradi_integrated_2025}, have been widely explored for operations such as matrix multiplication, data movement, and signal processing \cite{hamerly_large-scale_2019, zhou_photonic_2022, miller_device_2009, shen_deep_2017}, but often rely on electronic subsystems to implement the nonlinear transformations essential to deep learning. 

Driven--dissipative solitons in Kerr resonators are a physical system of discrete nonlinear excitations in an open resonator field, sustained by a coherent external drive \cite{leo_temporal_2010, tobiasreview2018}. The resulting out-coupled, ultrafast pulse train forms an optical frequency comb \cite{spencer_optical-frequency_2018, drake_terahertz-rate_2019, drake_thermal_2020}. Therefore, soliton microcombs have been explored and are being optimized as light sources for data communications \cite{pirmoradi_integrated_2025}, tensor multiplication for optical computing \cite{xu_11_2021, feldmann_parallel_2021}, wideband signal generation \cite{zang_universal_2025}, and ranging \cite{suh_soliton_2018, riemensberger_massively_2020}. Ensembles of temporally separated solitons are naturally stable \cite{cole_soliton_2017}, providing a platform in which nonlinear dynamics give rise to addressable states that can be controlled and multiplexed in time \cite{leo_temporal_2010}. In systems with synchronously pulsed or modulated driving, the external drive can additionally trap the solitons at well-defined temporal positions \cite{Obrzud2017Temporal,Hendry_trapping_2018,Cole_chirped_2018}. This capability has enabled quantum-optical computing using soliton dynamics under Kerr Hamiltonians, including a soliton Ising machine for combinatorial optimization \cite{jin_combinatorial_2026,chauhan2026tripartite}. 

\begin{figure*}[t]
    \centering
    \includegraphics[width=0.9\textwidth]{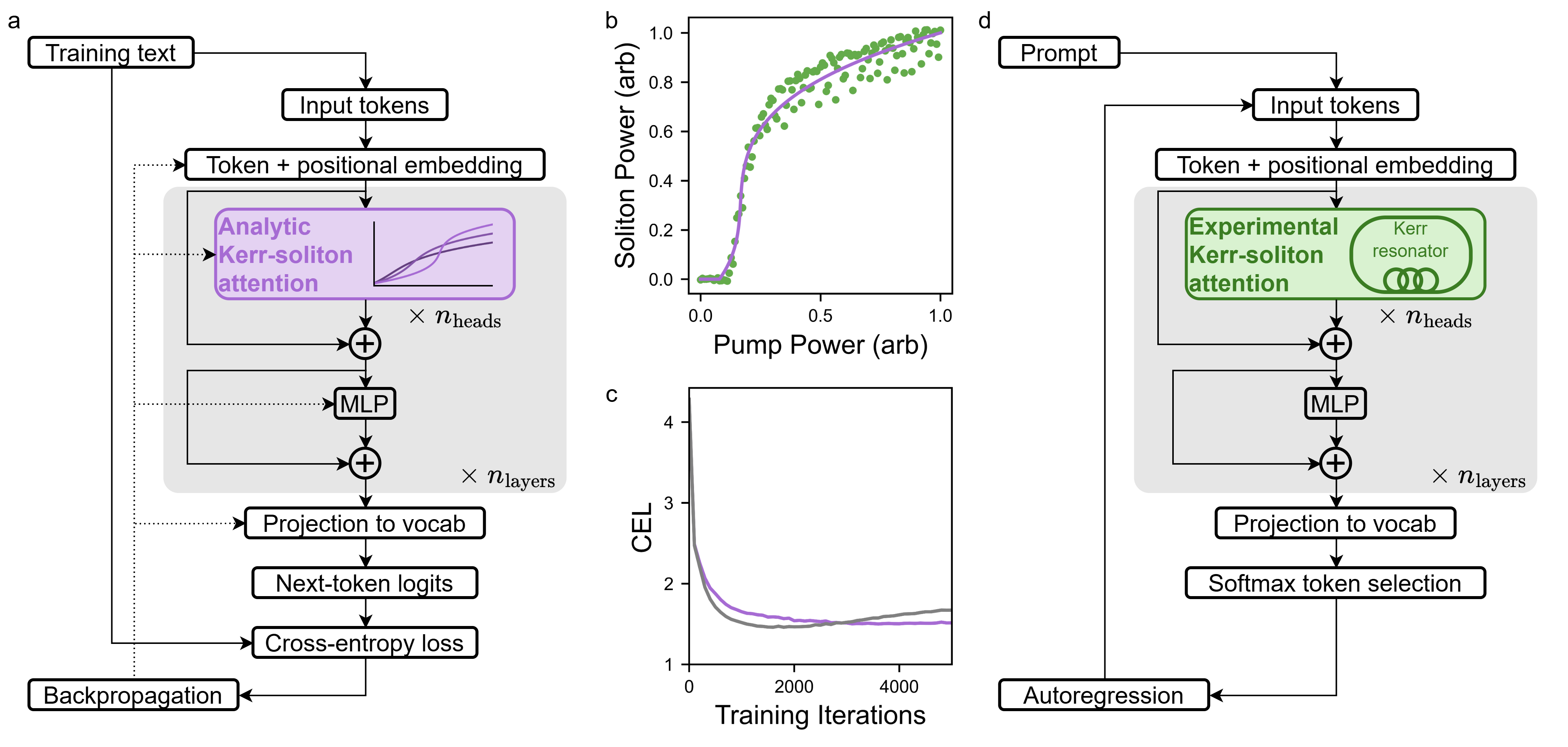}
    \caption{\textbf{Training and inference framework for Kerr-soliton attention.}
    (\textbf{a}) Deep-learning training framework for analytic Kerr-soliton attention. The attention layer of the deep-learning model features the analytic Kerr-soliton response of Eq.~\eqref{eq:cubic}. Gradient-based backpropagation updates model weights to complete the training loop. (\textbf{b}) Kerr-nonlinear resonator calibration. The measured pump-input-to-soliton-output relation (green) is fit to Eq.~\eqref{eq:cubic} (purple) with an effective detuning of $\alpha_{\mathrm{eff}}=1.49\pm0.09$. (\textbf{c}) Validation cross-entropy loss (CEL) of the analytic Kerr-soliton attention model (purple) and softmax attention model (gray) as a function of training iterations. (\textbf{d}) Framework for autoregressive inference using Kerr-soliton attention. An ensemble of Kerr solitons replaces the analytic model used in training. Model weights are the same as those optimized in training with the analytic model.}
    \label{fig:design}
\end{figure*}

Here, we leverage the driven--dissipative dynamics of an ensemble of Kerr solitons to provide computing for deep-learning models. We design a hybrid digital--physical generative pre-trained transformer (GPT) in which the soliton ensemble encodes token representations and implements nonlinear transformations through its intrinsic dynamics. We use gradient-based training to realize a language model with an analytic representation of Kerr-soliton attention, and we explore generative inference by deploying the trained model on soliton hardware. Token generation emerges from the nonlinear input--output relation of the Kerr soliton system, which we refer to as the soliton response. We experimentally benchmark the agreement between the analytic response used during training and the physical Kerr-soliton response under inference-generated inputs, finding a root mean square error in the normalized outputs of less than 0.01. We demonstrate operating the GPT with 256 solitons in flight, streaming model score data while the resulting nonlinear weight vector is stored and processed by the solitons.

Figure~\ref{fig:design} outlines a deep-learning GPT architecture based on the Kerr-soliton input--output relation as a physical nonlinear response for attention. Our experiments proceed in two phases: digital, gradient-based model training with an analytic representation of the soliton response and generative inference with nonlinear compute performed by the physical system. We illustrate the training process of our deep-learning model to perform computation with a physical nonlinear element; see Fig.~\ref{fig:design}(a). The model's input text is tokenized and used to generate queries ($Q \in \mathbb{R}^{L\times d_k}$), keys ($K \in \mathbb{R}^{L\times d_k}$), and values ($V \in \mathbb{R}^{L \times d_v}$), where $d_k$ and $d_v$ are the embedding sizes for a single head of attention for the $Q$, $K$, and $V$ matrices, respectively. The score matrix, $S = QK^T/\sqrt{d_k}$, encodes the relevance of tokens to each other, while a row of attention, $A = wV$, is a representation of what token $i$ has learned from the sequence, which uses the soliton response $Y$ to convert scores into nonlinear weights, $w=Y(S)$, and capture arbitrary sequence characteristics \cite{yun_are_2020}. As the soliton response, $Y$ is nonnegative and nonlinear, as are commonly used functions such as softmax \cite{vaswani_attention_2017}. To train the model to leverage $Y$, next-token predictions are compared against training text and the model is optimized using gradient-based backpropagation. In this hybrid digital--physical GPT, we perform analytic differentiation of $Y$ for training the deep-learning model,  enabling efficient optimization against the analytic response without experimental noise.

The driven--dissipative Kerr-soliton dynamics we use for deep learning are described by the Lugiato--Lefever equation (LLE), a nonlinear Schrödinger equation describing the mean field of a coherently driven resonator \cite{godey2014stability, cole_soliton_2017}. The normalized LLE has the form
\begin{equation}\label{eq:lle}
\frac{\partial\psi}{\partial\tau} = -(1+i\alpha)\psi
+ i\frac{d_2}{2}\frac{\partial^2 \psi}{\partial \theta^2}
+ i|\psi|^2 \psi + F(\theta),
\end{equation}
where $\psi$ is the intraresonator mean field, $\alpha$ is the normalized physical detuning, $d_2$ is the dispersion parameter, and $F(\theta)$ is the driving field, all with unit normalization (Methods). In the LLE, an ensemble of temporally separated solitons can arise in many configurations, especially in the case of a series of discrete pump pulses. The dynamics of the ensemble, mediated by the Kerr interaction and $F$, can also exhibit complex self-, cross-, and ensemble-wide interactions \cite{jin_combinatorial_2026,jin2026nanophotoniccontrolcollectivemanybody}. For the attention transformer, we encode a row of the score matrix across an ensemble of $L$ pump pulses as 
\begin{equation}\label{eq:pump}
    F_i(\theta) = \sum_{j=1}^{L}F_0\left(\theta + \frac{2\pi (j-1)}{L}\right)\sqrt{f\left(S_{i, j}\right)},
\end{equation}
where $f$ maps model scores to dimensionless relative pump powers, with negative scores mapped to zero and a temperature-like parameter setting the conversion between score amplitude and relative pump power, and $F_0(\theta)$ is the laser pump pulse waveform.  The form of $f$ depends on the specific experimental realization and is subject to stability requirements that can be estimated with the LLE.

To represent the Kerr-soliton response during digital training, we adopt the stationary input--output relation of a driven Kerr cavity as an analytic model (Methods). The response is defined through the implicit relation
\begin{equation}\label{eq:cubic}
Y^3 - 2\alpha_{\mathrm{eff}}Y^2
+ (1+\alpha_{\mathrm{eff}}^2)Y - X = 0,
\end{equation}
\noindent where $X$ and $Y$ are calibrated input and output coordinates corresponding to the programmed pump intensity and detected soliton power, respectively, and $\alpha_{\mathrm{eff}}$ controls the shape of the analytic response. Thus, $f(S_{i,j})$ sets the physical pump power, which is mapped by an experimental calibration onto the normalized input coordinate $X$; Eq.~\eqref{eq:cubic} then determines the corresponding nonlinear output $Y$.

We construct a deep-learning model around this nonlinear response, enabling direct transfer of trained parameters to the experimental system for inference. We use an ensemble of temporally separated Kerr solitons in a single resonator, driven by pump pulses whose intensities are set by a time-multiplexed electrical input applied to a modulator \cite{jin_combinatorial_2026}. To define $f$, we first calibrate the experimentally accessible pump-power range and its mapping to $X$, then fit the measured soliton response to determine $\alpha_{\mathrm{eff}}$. We plot soliton powers (green points) in Fig.~\ref{fig:design}(b) as a function of the input pump power and fit the data to Eq.~\eqref{eq:cubic} (purple line), obtaining an effective detuning of $\alpha_{\mathrm{eff}}=1.49\pm0.09$.

We train a transformer with Kerr-soliton attention, and we benchmark performance with respect to a standard softmax-attention model. We train both models on a character-level tokenization of the text of Shakespeare \cite{karpathy_shakespeare_2015} (Methods). We plot the validation cross-entropy loss (CEL) in Fig.~\ref{fig:design}(c) for the Kerr-soliton model in purple and for the softmax model in gray as a function of the number of training iterations. The Kerr-soliton model trains successfully without optimization failure, reaching a minimum CEL of 1.501 after 3600 iterations, compared with 1.461 after 1800 iterations for the softmax baseline. Importantly, the similarity in training behavior indicates that the analytic Kerr-soliton response can serve as a viable attention mechanism for deep learning.

\begin{figure*}[t]
    \centering
    \includegraphics[width=0.9\textwidth]{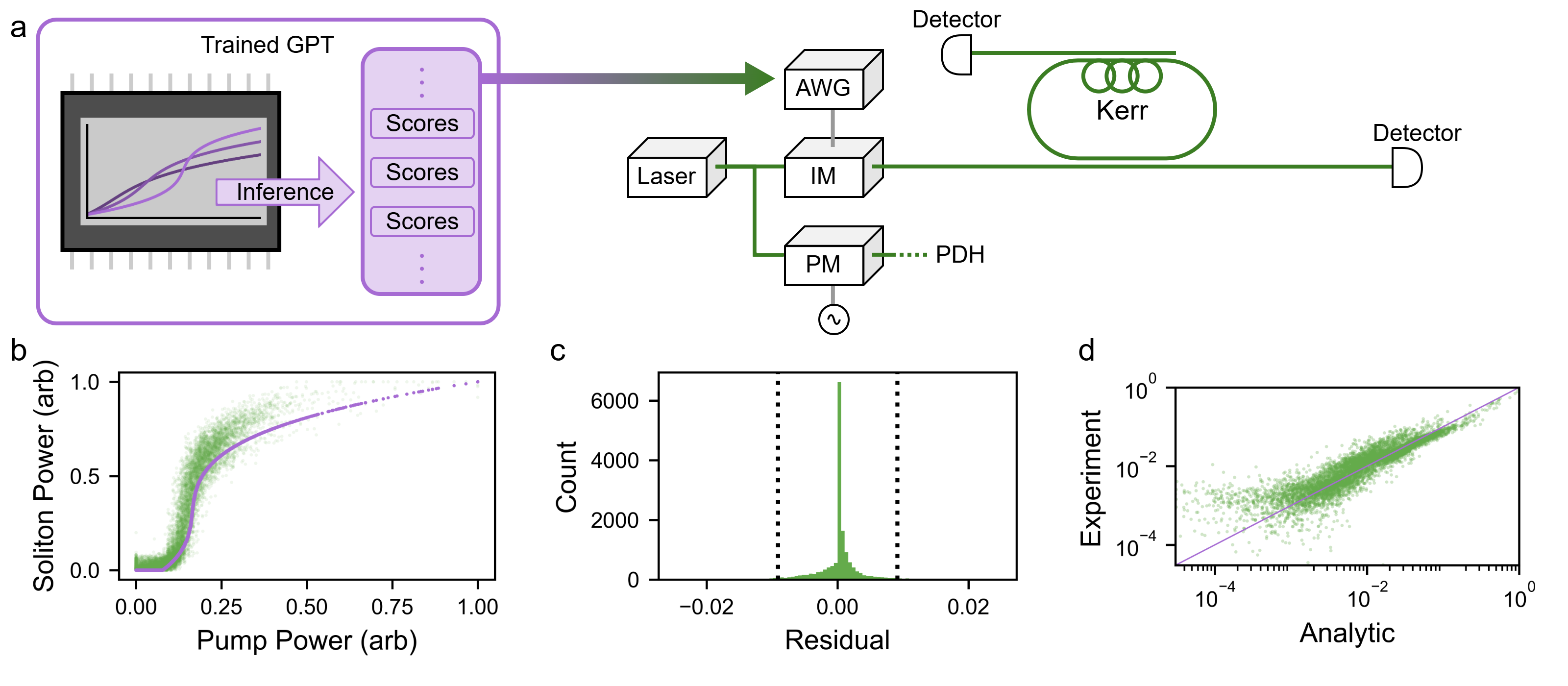}
    \caption{\textbf{Experimental validation of Kerr-soliton attention.}
    (\textbf{a}) System for generative inference experiments. Trained model weights are used to generate scores that are encoded on a pump laser using an intensity modulator (IM) driven by an arbitrary waveform generator (AWG). The laser is Pound--Drever--Hall (PDH) locked to the Kerr resonator through a phase-modulated (PM) path. Before detection, the center pump frequency is filtered out. (\textbf{b}) Normalized nonlinear weights calculated from experimental results plotted against the analytic Kerr-soliton relation. (\textbf{c}) Residual of nonlinear weights generated by experiment and the analytic relation. Vertical lines are shown at the RMSE of 0.0091. (\textbf{d}) Experimental nonlinear weights plotted against those computed using the analytic relation.}
    \label{fig:validation}
\end{figure*}

Figure~\ref{fig:design}(d) describes the generative-inference process in our experiments, using the trained deep-learning model with operation on experimental hardware. The model computes the final row of the score matrix and encodes the elements onto the soliton ensemble via Eq.~\eqref{eq:pump}. After nonlinear evolution, the measured soliton powers are digitized and mapped back into nonlinear weights. After multiplying by $V$, this completes the computation for the final row of attention, which is then processed to predict the next token. In autoregressive operation, this procedure is repeated as each generated token is appended to the model context. In this work, Kerr solitons provide the nonlinear attention response and time-multiplexed, dynamic memory of the attention vector, and suggest a pathway toward implementing additional nonlinear components, such as the multi-layer perceptron \cite{hamerly_large-scale_2019, zhou_photonic_2022, shen_deep_2017}. We use digital computing for token embedding, matrix multiplication, row normalization, and output projection, operations that are highly efficient in conventional electronics.

Figure~\ref{fig:validation} presents experimental validation of Kerr-soliton attention using data generated during autoregressive language inference with the trained model. We operate the trained GPT digitally using the analytic Kerr-soliton response and sample score vectors produced during language generation. These inference-generated scores are then encoded and replayed on the physical Kerr-soliton system, allowing us to compare the experimentally produced nonlinear weights directly with the analytic response used during training. The experimental system comprises a laser, intensity modulator, optical-fiber resonator, and photodetector; see Fig.~\ref{fig:validation}(a). The resonator consists of $25~\mathrm{m}$ of single-mode fiber with two external couplers and has a round-trip time of $T_r=138~\mathrm{ns}$. A CW laser ($\lambda=1550~\mathrm{nm}$) is frequency locked to the resonator using the Pound--Drever--Hall technique, while intensity and phase modulation clocked to a common microwave reference generate an electro-optic pulse train with $L$ pulses \cite{beha_electronic_2017}. The intensity of each pulse is individually controlled by an arbitrary waveform generator (AWG) according to Eq.~\eqref{eq:pump} within the operating range of intensities determined by $f$. We group causally masked inference scores into batches of $L=256$ for experimental processing, enabling inference-relevant hardware measurements without fully serialized autoregressive acquisition. After programming each pump pattern, we allow the solitons to settle and acquire repeated photodetection traces. We take the median across these measurements to estimate the soliton power in each time slot, which is then mapped to nonlinear weights for subsequent digital processing.

By measuring the soliton powers produced by inference-generated pump inputs, we test the calibrated Kerr response under the operating conditions encountered by the trained model. Figure~\ref{fig:validation}(b) presents the measured soliton power as a function of input pump power together with the analytic Kerr-soliton response used during training. Importantly, this measurement tests the response across the distribution of inputs produced during language inference and across the time-multiplexed soliton ensemble, rather than only along the calibration sweep of Fig.~\ref{fig:design}(b). After row normalization, the residuals between the experimental and analytic nonlinear weights have a root-mean-square error of 0.0091; see Fig.~\ref{fig:validation}(c). A direct comparison of the experimental and analytic weights in Fig.~\ref{fig:validation}(d) clusters around the diagonal, demonstrating that the calibrated nonlinear response is preserved across the full soliton ensemble and over more than 20 dB of output dynamic range. Together, these measurements demonstrate a continuous, uniform, and stable nonlinear mapping from pump power to soliton power. Moreover, because the soliton is an intrinsic nonlinear eigenstate of the Kerr resonator, its  response seamlessly interfaces to digital computation.

\begin{figure*}[t]
    \centering
    \includegraphics[width=0.9\textwidth]{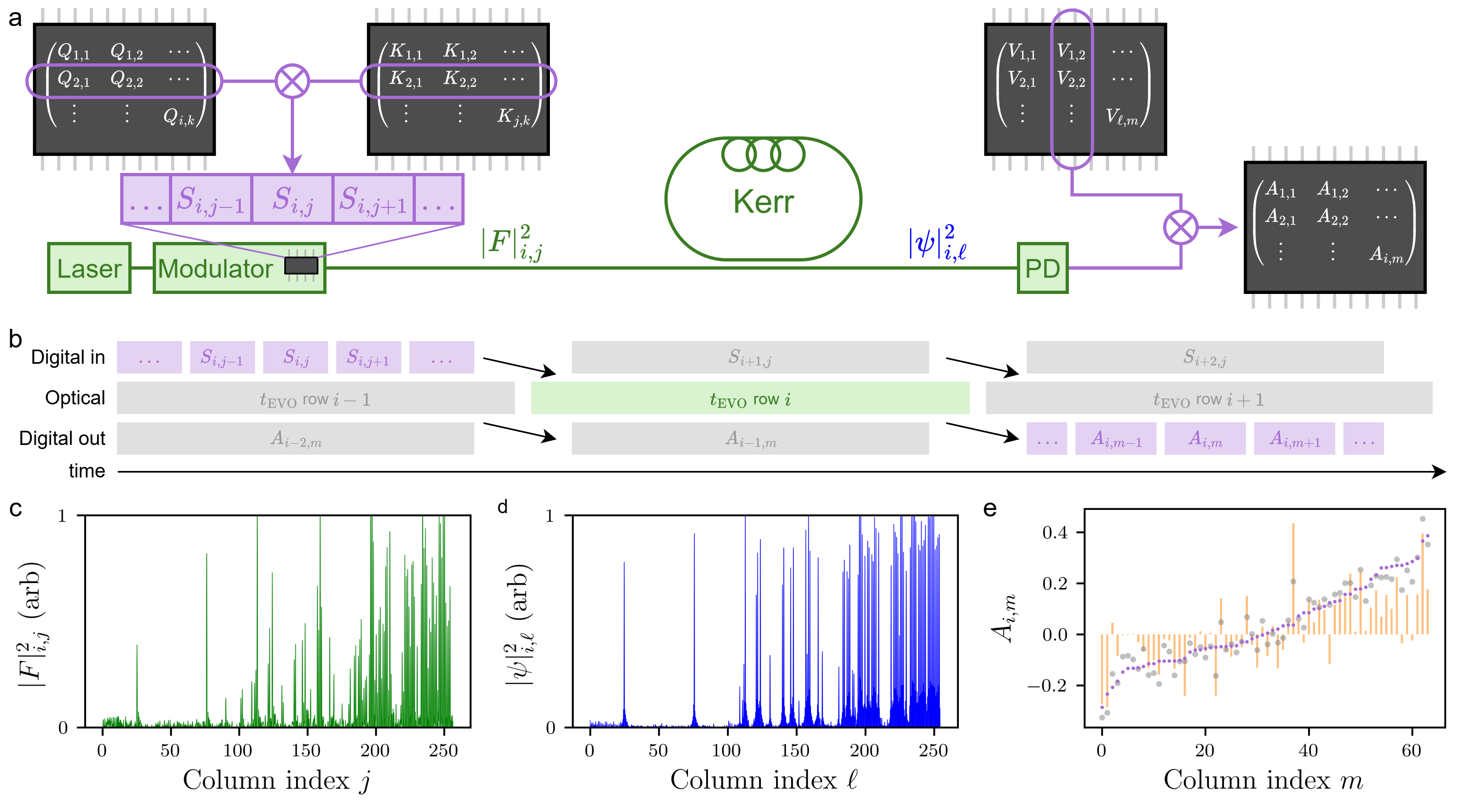}
    \caption{\textbf{Streamed Kerr-soliton attention.}
    (\textbf{a}) Calculation for one row of the score matrix. Model scores, $S_{i,j}$, are calculated from queries ($Q$) and keys ($K$) and streamed to the intensities of an ensemble of pump pulses using an optical modulator. After photodetection, the experimentally calculated nonlinear weights are combined with the value matrix to produce a row of attention, $A_{i,m}$. (\textbf{b}) Timeline of streamed operation. While row $i$ evolves optically for a time $t_{\mathrm{EVO}}$, digital inputs for the subsequent row and nonlinear weight outputs from the previous row can be prepared simultaneously. (\textbf{c}) Measured pump waveform $|F_{i,j}|^2$ encoding the streamed row of the score matrix. (\textbf{d}) Measured soliton powers $|\psi_{i,\ell}|^2$ after evolution for $t_{\mathrm{EVO}}$. (\textbf{e}) Resulting row of attention, $A_{i,m}$, from the experimental nonlinear weights (purple), compared with the analytic Kerr-soliton response (gray) and the corresponding digital softmax calculation (orange). The column index $m$ is sorted based on the value of the experimental attention result.}
    \label{fig:streaming}
\end{figure*}

Figure~\ref{fig:streaming} explores high-speed, streamed generative inference with the attention weights stored and processed within the soliton ensemble. In this experiment, we analyze the soliton response in real-time computation without excess evolution time or averaging over multiple photodetections. Using the driven--dissipative Kerr dynamics, we motivate a minimum evolution time based on the system's photon lifetime, operating power, and desired measurement precision. We outline the procedure for full row-wise attention in the hybrid model; see Fig.~\ref{fig:streaming}(a). The selected $Q$ row vector is digitally multiplied by the $K$ matrix to form one row of the score matrix, $S_{i,j}$. We stream the scores to the intensity of pump pulses, encoding them across a time-multiplexed ensemble with the number of pulses matching the context length of the trained model, which we set to 256 in this work. The soliton ensemble evolves these scores in parallel, and the resulting row of nonlinear weights is combined with the value matrix to form the row of attention, $A_{i,m}$. The conversion from scores to nonlinear weights occurs directly through evolution of the soliton ensemble, which can proceed alongside digital input preparation and output processing.

To analyze the trade-off between nonlinear weight processing speed and power consumption in streamed Kerr-soliton attention, we consider the soliton evolution time, $t_{\mathrm{EVO}}$, which is adjustable by the system parameters. We define $t_{\mathrm{EVO}}$ as the time required for the soliton power to approach steady state within the bit depth of an $x$-bit measurement (Methods). For our system parameters, we calculate the evolution time to be $t_{\mathrm{EVO}} \approx 1~\mu\mathrm{s}$, which is consistent with our previous Kerr-soliton combinatorial optimization experiments \cite{jin_combinatorial_2026, chauhan2026tripartite}. In the streamed experiment, we use an evolution time of $20T_r=2.76~\mu\mathrm{s}$, providing a conservative evolution window relative to this estimate. Increasing the external coupling and effective pump power can decrease $t_{\mathrm{EVO}}$ at an equivalent nonlinear operating point (Methods).

We experimentally demonstrate the streamed execution of a complete row of Kerr-soliton attention and quantify its operating rate. Figure~\ref{fig:streaming}(b) illustrates the timing of streamed operation, in which digital preparation of the subsequent score row and post-processing of the previous result can proceed during the optical evolution of the current row. Because the evolution time is independent of the number of solitons in the time-multiplexed ensemble, increasing the context length increases the aggregate score-processing rate. Fig.~\ref{fig:streaming}(c)--(e) shows the corresponding experimental calculation: the modulated pump waveform $|F_{i,j}|^2$ encodes row $i$ of the score matrix, the measured soliton powers $|\psi_{i,\ell}|^2$ provide the nonlinear weights after evolution for $t_{\mathrm{EVO}}$, and these weights are combined with $V$ to produce the attention output $A_{i,m}$. The resulting attention row agrees closely with the analytic Kerr-soliton calculation and is shown alongside the corresponding digital softmax result. The streamed measurement is performed without averaging over repeated photodetections and corresponds to an aggregate processing rate of approximately $9\times10^7$ score elements per second, or approximately $0.7~\mathrm{Gb/s}$ at 8-bit resolution.

\begin{figure}[t]
    \centering
    \includegraphics[width=0.8\linewidth]{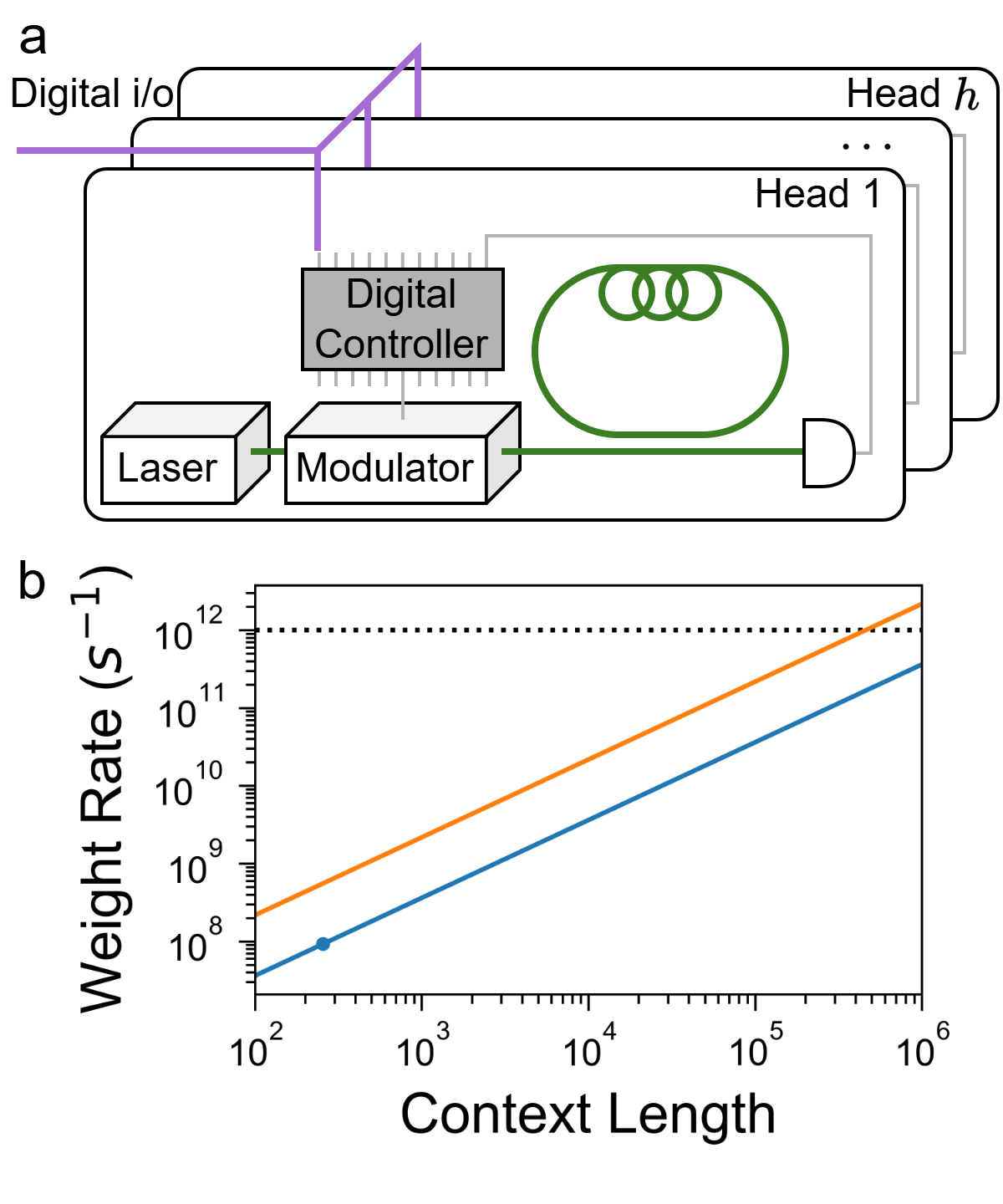}
    \caption{\textbf{Scaling Kerr-soliton attention.}
    (\textbf{a}) Diagram for parallel heads using Kerr-soliton attention. A shared digital path streams model scores to digital controllers that manage each head independently. (\textbf{b}) Estimated weight-calculation rate for a single head (blue line), including the experimental streamed calculation (blue dot), and six parallel heads (orange line) compared to an estimate for digital weight calculation limited by memory bandwidth (black dashed line).}
    \label{fig:scaling}
\end{figure}

Since modern GPT language models involve token context lengths greater than $10^5$, we assess the prospects for scaling Kerr-soliton attention in Fig.~\ref{fig:scaling}, while leveraging the key advantage of digital training and hardware inference. In optical-spectrum-analyzer measurements of the driven solitons, we observe a spectral bandwidth of approximately $3~\mathrm{nm}$; for a transform-limited $\mathrm{sech}^2$ soliton, this corresponds to a duration of approximately $0.8~\mathrm{ps}$. Since the soliton pulse duration is governed principally by the resonator dispersion and detuning rather than pump-pulse duration \cite{Obrzud2017Temporal, Lucas_detuning_2017}, shorter pump pulses can support a denser ensemble of otherwise similar solitons. For the $138~\mathrm{ns}$ round-trip time used here, the transform-limited soliton duration gives a packing limit of approximately $10^5$ solitons. The soliton ensemble also supports parallelization through independent attention heads that leverage a shared $K$ and $V$ data stream; see Fig.~\ref{fig:scaling}(a). The memory stream distributes model data to multiple controlled photonic circuits, each comprising the laser, modulation, resonator, and detection elements. The digital controller associated with each Kerr-soliton attention head provides the interface between the trained model and the photonic system, setting the score encoding, timing, normalization, and operating condition.

We consider weight generation scaling as a function of $L$; see Fig.~\ref{fig:scaling}(b). Here, we assume a representative $1~\mathrm{TB/s}$ digital memory bandwidth \cite{nvidia_a100_datasheet_2021}, equivalent to a nonlinear weight-generation rate of $10^{12}$ weights per second for 8-bit encoded weights. For one attention head with $L$ solitons, the nonlinear weight-computation rate scales as $L/t_{\mathrm{EVO}}$, since $L$ score elements evolve during the same physical evolution window, and with $H$ parallel heads, the rate scales as $HL/t_{\mathrm{EVO}}$. In Fig.~\ref{fig:scaling}(b), we plot the weight-computation rate and the memory-bandwidth limit (black dashed line) for two Kerr-soliton GPT configurations, $H=1$ (blue line) and $H=6$ (orange line); we include the experimental operating point of our streamed computation in the same plot (blue dot). At the soliton-packing limit estimated above, multiple parallel heads approach the representative digital memory-bandwidth scale, demonstrating that shorter pump pulses and parallel photonic channels can increase the aggregate nonlinear weight-generation rate.

The parallelization structure in Fig.~\ref{fig:scaling}(a) additionally indicates a broader design space for driven--dissipative dynamics in Kerr-soliton GPTs. First, real-time control of the system hyperparameters, e.g. $f(S)$, $\alpha$, and $t_{\mathrm{EVO}}$, enables a host of optimization opportunities. These include pre-characterizing $Y(S)$ to select the most relevant portions of $V$ to perform efficient computation of $A$, use of transient Kerr dynamics to estimate $Y(S)$ prior to steady state, and tuning $\alpha$ to control selectivity and sparsity of the response. Second, in a multi-head Kerr-soliton GPT, each physical channel introduces an additional degree of freedom, as individual heads can be configured through their optical encoding and resonator operating conditions, enabling head-dependent nonlinear transformations while remaining compatible with a shared digitally trained model and a common analytic--experimental mapping for inference.

In conclusion, we have introduced and experimentally validated a hybrid digital--physical transformer in which Kerr-soliton dynamics implement the nonlinear attention operation. By training the model using an analytic Kerr-soliton response, we enable gradient-based optimization that is matched to the physical system. We demonstrate that nonlinear attention weights are generated through real-time evolution of a time-multiplexed soliton ensemble, with high-fidelity agreement between analytic predictions and experimental measurements. This approach enables streaming-in-time computation in which memory and nonlinear processing are co-located in the physical dynamics, reducing intermediate data movement. The soliton system therefore acts as a high-bandwidth nonlinear processor embedded within a deep-learning model. More broadly, this work establishes analytic training with physical inference as a viable paradigm for integrating photonic systems into modern machine learning architectures.

\section*{Methods}\label{sec:Methods}

\subsection*{LLE normalization}

In Eq.~\eqref{eq:lle}, we use a normalized mean-field Lugiato--Lefever equation for the intracavity field.
A dimensional form consistent with Eq.~\eqref{eq:lle} is
\begin{equation}\label{eq:lle_dim}
\begin{aligned}
\frac{\partial A(t,\theta)}{\partial t}
&=\left[-\frac{\kappa}{2}-i(\omega_0-\omega_p)\right.\\
&\qquad\left.+i\frac{D_2}{2}\frac{\partial^2}{\partial\theta^2}
+i g|A|^2\right]A\\
&\quad+\sqrt{\kappa_{\mathrm{ex}}}\,S(\theta),
\end{aligned}
\end{equation}
where $A(t,\theta)$ is the dimensional intracavity field, $t$ is the slow physical time, $\theta$ is the angular coordinate around the resonator, $\kappa=\kappa_0+\kappa_{\mathrm{ex}}$ is the total cavity linewidth, $\kappa_0$ and $\kappa_{\mathrm{ex}}$ are the intrinsic and coupling loss rates, $\omega_0$ is the cavity resonance frequency, $\omega_p$ is the pump frequency, $D_2$ is the second-order integrated dispersion coefficient, $g$ is the dimensional single-photon Kerr shift, and $S(\theta)$ is the drive field.
Here we take $|A|^2$ to denote intracavity photon number and $|S|^2$ to denote incident photon flux.

The normalized variables are defined by
\begin{equation}\label{eq:fnorm}
\tau=\frac{\kappa t}{2},
\qquad
\psi=\sqrt{\frac{2g}{\kappa}}\,A,
\qquad
F(\theta)=\sqrt{\frac{8g\kappa_{\mathrm{ex}}}{\kappa^3}}\,S(\theta),
\end{equation}
together with the normalized detuning and dispersion parameters
\begin{equation}
\alpha=\frac{2(\omega_0-\omega_p)}{\kappa},
\qquad
d_2=\frac{2D_2}{\kappa}.
\end{equation}
With these definitions, Eq.~\eqref{eq:lle_dim} reduces to Eq.~\eqref{eq:lle}.
\\

\subsection*{Analytic Kerr response}

Training the deep-learning model requires an analytic representation of the measured nonlinear response. We adopt the stationary input--output relation of a driven Kerr cavity as the functional form of this response.

For a normalized scalar cavity field $a$ driven by a field $s$,
\begin{equation}\label{eq:scalar_kerr}
\frac{da}{d\tau}
=
-(1+i\alpha_{\mathrm{eff}})a
+i|a|^2a
+s,
\end{equation}
where $\alpha_{\mathrm{eff}}$ is an effective detuning parameter. At stationary state,
\begin{equation}
s
=
\left[
1+i\left(
\alpha_{\mathrm{eff}}-|a|^2
\right)
\right]a.
\end{equation}
Defining
\begin{equation}
X=|s|^2,
\qquad
Y=|a|^2,
\end{equation}
gives the Kerr-cavity input--output relation
\begin{equation}\label{eq:kerr_response}
X
=
Y\left[
1+
\left(
\alpha_{\mathrm{eff}}-Y
\right)^2
\right].
\end{equation}
Expanding Eq.~\eqref{eq:kerr_response} gives
\begin{equation}
Y^3
-2\alpha_{\mathrm{eff}}Y^2
+
\left(
1+\alpha_{\mathrm{eff}}^2
\right)Y
-X
=0,
\end{equation}
which is Eq.~\eqref{eq:cubic}.

For the experimentally measured response, $X$ and $Y$ are calibrated input and output coordinates corresponding to the programmed pump intensity and detected soliton power, respectively. The parameter $\alpha_{\mathrm{eff}}$ determines the shape of the analytic response and is obtained from the experimental calibration. It is distinct from the normalized physical detuning $\alpha$ appearing in Eq.~\eqref{eq:lle}.
\\

\subsection*{Experimental response calibration}

We calibrate the analytic response using measurements of detected soliton power as a function of programmed pump intensity. Static input and output calibration parameters map these experimental quantities onto the normalized coordinates $X$ and $Y$ of Eq.~\eqref{eq:kerr_response}. We fit Eq.~\eqref{eq:cubic} to the measured response shown in Fig.~\ref{fig:design}(b), obtaining $\alpha_{\mathrm{eff}}=1.49\pm0.09$.

The calibrated operating range defines the interval of pump intensities used for score encoding. The function $f$ in Eq.~\eqref{eq:pump} maps model scores onto this interval. During experimental operation, the same calibration maps the programmed pump intensities onto $X$ and the measured soliton powers onto the normalized nonlinear weights used in the attention calculation.
\\

\subsection*{GPT model parameters}

For our Kerr-soliton and softmax GPTs, we use the publicly available nanoGPT by Andrej Karpathy \cite{karpathy_nanogpt_2022} as a foundation. We add the ability to switch between softmax and analytic Kerr-soliton attention while keeping other model parameters constant. For Kerr-soliton attention, scores in each row are mean-centered, divided by a temperature parameter of 5, and clipped to the interval $[-2,2]$. The resulting coordinate is scaled and evaluated using the analytic Kerr response of Eq.~\eqref{eq:cubic} with $\alpha_{\mathrm{eff}}=1.48$, with negative input coordinates mapped to zero; the resulting nonlinear weights are then normalized along each row. We evaluate this response using an 8192-point lookup table to expedite computation.

We train both models on a character-level tokenization of the text of Shakespeare \cite{karpathy_shakespeare_2015}, using backpropagation to compute gradients and AdamW for optimization. The cross-entropy loss (CEL) quantifies the negative log-likelihood of the correct next-character prediction under the model distribution, with lower values indicating improved predictive accuracy. We estimate CEL by averaging over 200 batches of next-character predictions randomly sampled from a reserved validation split of the data, and evaluate this estimate every 100 training iterations.

See Table~\ref{tab:params} for model parameters relevant to both Kerr-soliton and softmax versions of attention. \\

\begin{table}[h!]
    \centering
    \caption{\textbf{Parameters shared by the Kerr-soliton and softmax GPT models.}}
    \label{tab:params}
    \begin{tabular}{c|c}
       Parameter  & Value \\
         \hline
        Vocab size & 65\\
        $n_\text{heads}$ & 6\\
        $n_\text{layers}$ & 6\\
        $n_\text{embed}$ & 384\\
        Context length & 256\\
        Learning rate & $10^{-3}$\\
        Dropout fraction & 0.2\\

    \end{tabular}
\end{table}

\subsection*{Cavity evolution time}

The nonlinear response is read after the intracavity field has evolved toward its steady operating state. We estimate the characteristic evolution time from the field relaxation rate of the driven cavity. For a first-order cavity response with total linewidth $\kappa$, the field approaches its stationary value $A_{\mathrm{s}}$ from an initially empty cavity according to
\begin{equation}
A(t)
=
A_{\mathrm{s}}
\left(
1-e^{-\kappa t/2}
\right).
\end{equation}
The corresponding power is
\begin{equation}
P(t)
=
P_{\mathrm{s}}
\left(
1-e^{-\kappa t/2}
\right)^2.
\end{equation}
Defining the fractional difference from stationary power as
\begin{equation}
\epsilon
=
\frac{P_{\mathrm{s}}-P(t)}{P_{\mathrm{s}}},
\end{equation}
the time required to reach a specified $\epsilon$ is
\begin{equation}\label{eq:evo_general}
t(\epsilon)
=
\frac{2}{\kappa}
\ln\left[
\frac{1}{
1-\sqrt{1-\epsilon}
}
\right].
\end{equation}
For $\epsilon\ll1$,
\begin{equation}
t(\epsilon)
\approx
\frac{2}{\kappa}
\ln\left(
\frac{2}{\epsilon}
\right).
\end{equation}

For an $x$-bit representation, we use $\epsilon=2^{-x}$ as the target fractional power precision. This gives
\begin{equation}\label{eq:evo_time}
t_{\mathrm{EVO}}
\approx
\frac{2(x+1)}{\kappa}\ln 2.
\end{equation}

The dependence of this timescale on optical power follows from the normalized LLE drive. At the peak of a pump pulse, Eq.~\eqref{eq:fnorm} gives
\begin{equation}\label{eq:normalized_peak_drive}
|F_{\mathrm{pk}}|^2
=
\frac{
8g\kappa_{\mathrm{ex}}P_{\mathrm{pk}}
}{
\hbar\omega_0\kappa^3
}
=
\frac{
8g\eta P_{\mathrm{pk}}
}{
\hbar\omega_0\kappa^2
},
\end{equation}
where $P_{\mathrm{pk}}$ is the incident pump-pulse peak power and
\begin{equation}
\eta=\frac{\kappa_{\mathrm{ex}}}{\kappa}
\end{equation}
is the coupling fraction. For a fixed normalized operating point, $|F_{\mathrm{pk}}|$ is constant, giving
\begin{equation}
\kappa
=
\frac{1}{|F_{\mathrm{pk}}|}
\sqrt{
\frac{
8g\eta P_{\mathrm{pk}}
}{
\hbar\omega_0
}
}.
\end{equation}
Substitution into Eq.~\eqref{eq:evo_time} gives
\begin{equation}\label{eq:evo_time_power_general}
t_{\mathrm{EVO}}
=
2(x+1)\ln 2\,
|F_{\mathrm{pk}}|
\sqrt{
\frac{
\hbar\omega_0
}{
8g\eta P_{\mathrm{pk}}
}
}.
\end{equation}
For $|F_{\mathrm{pk}}|\sim1$,
\begin{equation}\label{eq:evo_time_power}
t_{\mathrm{EVO}}
\sim
2(x+1)\ln 2
\sqrt{
\frac{
\hbar\omega_0
}{
8g\eta P_{\mathrm{pk}}
}
}.
\end{equation}

For the experimental system and operating conditions used here, this estimate gives $t_{\mathrm{EVO}}\approx1~\mu\mathrm{s}$. The streamed acquisition contains five zero-pump buffer cycles followed by 20 pumped round trips. Only the pumped interval is included in the demonstrated evolution time, giving $t_{\mathrm{EVO}}=20T_r=2.76~\mu\mathrm{s}$.\\

\subsection*{AI-assisted code development}

OpenAI ChatGPT (GPT-4o through GPT-5.6) was used to assist
with code generation, revision, and debugging. The authors
specified the computational methods and requirements,
reviewed and revised the generated code, and verified the
resulting analyses against the underlying data and
calculations.

\begin{acknowledgments}
\paragraph*{Funding:} 
This research has been funded by AFOSR grant FA9550-20-1-0004 (Project 19RT1019),
NSF Quantum Leap Challenge Institute Award OMA-2016244, and NIST.
\end{acknowledgments}

\paragraph*{Author contributions:}
Conceptualization: L.M.W., Y.J., S.B.P. Experimental and computational work, data analysis, and visualization: L.M.W. Writing: L.M.W., with review and editing by all authors. Supervision, project administration, and funding acquisition: S.B.P.

\paragraph*{Competing interests:}
The authors declare no competing interests. This work is a contribution of the US Government and is not subject to US copyright. Mention of specific companies or trade names is for scientific communication only and does not constitute an endorsement by NIST.

\paragraph*{Data and code availability:}
The data and code supporting the findings of this
article are available from the corresponding author upon
reasonable request.

\bibliography{sbib_nourl}

@article{cole_soliton_2017,
  author = {Cole, Daniel C. and Lamb, Erin S. and Del'Haye, Pascal and Diddams, Scott A. and Papp, Scott B.},
  title = {Soliton crystals in {Kerr} resonators},
  journal = {Nature Photonics},
  volume = {11},
  number = {10},
  pages = {671--676},
  year = {2017},
  doi = {10.1038/s41566-017-0009-z},
}

@article{leo_temporal_2010,
  author = {Leo, Fran{\c c}ois and Coen, St{\'e}phane and Kockaert, Pascal and Gorza, Simon-Pierre and Emplit, Philippe and Haelterman, Marc},
  title = {Temporal cavity solitons in one-dimensional {Kerr} media as bits in an all-optical buffer},
  journal = {Nature Photonics},
  volume = {4},
  number = {7},
  pages = {471--476},
  year = {2010},
  doi = {10.1038/nphoton.2010.120},
}

@article{drake_terahertz-rate_2019,
  author = {Drake, Tara E. and Briles, Travis C. and Stone, Jordan R. and Spencer, Daryl T. and Carlson, David R. and Hickstein, Daniel D. and Li, Qing and Westly, Daron and Srinivasan, Kartik and Diddams, Scott A. and Papp, Scott B.},
  title = {Terahertz-{Rate} {Kerr}-{Microresonator} {Optical} {Clockwork}},
  journal = {Physical Review X},
  volume = {9},
  number = {3},
  pages = {031023},
  year = {2019},
  doi = {10.1103/PhysRevX.9.031023},
}

@article{drake_thermal_2020,
  author = {Drake, Tara E. and Stone, Jordan R. and Briles, Travis C. and Papp, Scott B.},
  title = {Thermal decoherence and laser cooling of {Kerr} microresonator solitons},
  journal = {Nature Photonics},
  volume = {14},
  number = {8},
  pages = {480--485},
  year = {2020},
  doi = {10.1038/s41566-020-0651-8},
  note = {Number: 8},
}

@article{pirmoradi_integrated_2025,
  author = {Pirmoradi, Ali and Zang, Jizhao and Omirzakhov, Kaisarbek and Yu, Zhehao and Jin, Yan and Papp, Scott B. and Aflatouni, Firooz},
  title = {Integrated multi-port multi-wavelength coherent optical source for beyond {Tb}/s optical links},
  journal = {Nature Communications},
  volume = {16},
  number = {1},
  pages = {6387},
  year = {2025},
  doi = {10.1038/s41467-025-61288-x},
}

@misc{zang_universal_2025,
  author = {Jizhao Zang and Travis C. Briles and Jesse S. Morgan and Andreas Beling and Scott B. Papp},
  title = {Universal electronic synthesis by microresonator-soliton photomixing},
  year = {2025},
  doi = {10.48550/arXiv.2505.08707},
  eprint = {2505.08707},
  archivePrefix = {arXiv},
  primaryClass = {physics.optics},
  note = {arXiv:2505.08707},
}

@article{momeni_training_2025,
  author = {Momeni, Ali and Rahmani, Babak and Scellier, Benjamin and Wright, Logan G. and McMahon, Peter L. and Wanjura, Clara C. and Li, Yuhang and Skalli, Anas and Berloff, Natalia G. and Onodera, Tatsuhiro and Oguz, Ilker and Morichetti, Francesco and del Hougne, Philipp and Le Gallo, Manuel and Sebastian, Abu and Mirhoseini, Azalia and Zhang, Cheng and Markovi{\'c}, Danijela and Brunner, Daniel and Moser, Christophe and Gigan, Sylvain and Marquardt, Florian and Ozcan, Aydogan and Grollier, Julie and Liu, Andrea J. and Psaltis, Demetri and Al{\`u}, Andrea and Fleury, Romain},
  title = {Training of physical neural networks},
  journal = {Nature},
  volume = {645},
  number = {8079},
  pages = {53--61},
  year = {2025},
  doi = {10.1038/s41586-025-09384-2},
}

@article{xu_11_2021,
  author = {Xu, Xingyuan and Tan, Mengxi and Corcoran, Bill and Wu, Jiayang and Boes, Andreas and Nguyen, Thach G. and Chu, Sai T. and Little, Brent E. and Hicks, Damien G. and Morandotti, Roberto and Mitchell, Arnan and Moss, David J.},
  title = {11 {TOPS} photonic convolutional accelerator for optical neural networks},
  journal = {Nature},
  volume = {589},
  number = {7840},
  pages = {44--51},
  year = {2021},
  doi = {10.1038/s41586-020-03063-0},
}

@article{mohseni2022ising,
  author = {Mohseni, Naeimeh and McMahon, Peter L. and Byrnes, Tim},
  title = {Ising machines as hardware solvers of combinatorial optimization problems},
  journal = {Nature Reviews Physics},
  volume = {4},
  number = {6},
  pages = {363--379},
  year = {2022},
  doi = {10.1038/s42254-022-00440-8},
}

@article{tobiasreview2018,
  author = {Kippenberg, Tobias J. and Gaeta, Alexander L. and Lipson, Michal and Gorodetsky, Michael L.},
  title = {Dissipative Kerr solitons in optical microresonators},
  journal = {Science},
  volume = {361},
  number = {6402},
  pages = {eaan8083},
  year = {2018},
  doi = {10.1126/science.aan8083},
}

@article{inagaki2016,
  author = {Inagaki, Takahiro and Haribara, Yoshitaka and Igarashi, Koji and Sonobe, Tomohiro and Tamate, Shuhei and Honjo, Toshimori and Marandi, Alireza and McMahon, Peter L. and Umeki, Takeshi and Enbutsu, Koji and Tadanaga, Osamu and Takenouchi, Hirokazu and Aihara, Kazuyuki and Kawarabayashi, Ken-ichi and Inoue, Kyo and Utsunomiya, Shoko and Takesue, Hiroki},
  title = {A coherent Ising machine for 2000-node optimization problems},
  journal = {Science},
  volume = {354},
  number = {6312},
  pages = {603-606},
  year = {2016},
  doi = {10.1126/science.aah4243},
}

@article{feldmann_parallel_2021,
  author = {Feldmann, J. and Youngblood, N. and Karpov, M. and Gehring, H. and Li, X. and Stappers, M. and Le Gallo, M. and Fu, X. and Lukashchuk, A. and Raja, A. S. and Liu, J. and Wright, C. D. and Sebastian, A. and Kippenberg, T. J. and Pernice, W. H. P. and Bhaskaran, H.},
  title = {Parallel convolutional processing using an integrated photonic tensor core},
  journal = {Nature},
  volume = {589},
  number = {7840},
  pages = {52--58},
  year = {2021},
  doi = {10.1038/s41586-020-03070-1},
}

@article{godey2014stability,
  author = {Godey, Cyril and Balakireva, Irina V and Coillet, Aur{\'e}lien and Chembo, Yanne K},
  title = {Stability analysis of the spatiotemporal Lugiato-Lefever model for Kerr optical frequency combs in the anomalous and normal dispersion regimes},
  journal = {Physical Review A},
  volume = {89},
  number = {6},
  pages = {063814},
  year = {2014},
  doi = {10.1103/PhysRevA.89.063814},
}

@article{beha_electronic_2017,
  author = {Beha, Katja and Cole, Daniel C. and Del'Haye, Pascal and Coillet, Aur{\'e}lien and Diddams, Scott A. and Papp, Scott B.},
  title = {Electronic synthesis of light},
  journal = {Optica},
  volume = {4},
  number = {4},
  pages = {406--411},
  year = {2017},
  doi = {10.1364/OPTICA.4.000406},
}

@article{spencer_optical-frequency_2018,
  author = {Spencer, D. T. and Drake, T. and Briles, T. C. and Stone, J. and Sinclair, L. C. and Fredrick, C. and Li, Q. and Westly, D. and Ilic, B. R. and Bluestone, A. and Volet, N. and Komljenovic, T. and Chang, L. and Lee, S. H. and Oh, D. Y. and Suh, M. G. and Yang, K. Y. and Pfeiffer, M. H. P. and Kippenberg, T. J. and Norberg, E. and Theogarajan, L. and Vahala, K. and Newbury, N. R. and Srinivasan, K. and Bowers, J. E. and Diddams, S. A. and Papp, S. B.},
  title = {An optical-frequency synthesizer using integrated photonics},
  journal = {Nature},
  volume = {557},
  number = {7703},
  pages = {81--85},
  year = {2018},
  doi = {10.1038/s41586-018-0065-7},
}

@inproceedings{vaswani_attention_2017,
  author = {Vaswani, Ashish and Shazeer, Noam and Parmar, Niki and Uszkoreit, Jakob and Jones, Llion and Gomez, Aidan N and Kaiser, {\L} ukasz and Polosukhin, Illia},
  title = {Attention is {All} you {Need}},
  booktitle = {Advances in {Neural} {Information} {Processing} {Systems}},
  volume = {30},
  publisher = {Curran Associates, Inc.},
  year = {2017},
}

@misc{kaplan_scaling_2020,
  author = {Kaplan, Jared and McCandlish, Sam and Henighan, Tom and Brown, Tom B. and Chess, Benjamin and Child, Rewon and Gray, Scott and Radford, Alec and Wu, Jeffrey and Amodei, Dario},
  title = {Scaling {Laws} for {Neural} {Language} {Models}},
  year = {2020},
  doi = {10.48550/arXiv.2001.08361},
  eprint = {2001.08361},
  archivePrefix = {arXiv},
  primaryClass = {cs.LG},
  note = {arXiv:2001.08361},
}

@inproceedings{brown_language_2020,
  author = {Brown, Tom and Mann, Benjamin and Ryder, Nick and Subbiah, Melanie and Kaplan, Jared D and Dhariwal, Prafulla and Neelakantan, Arvind and Shyam, Pranav and Sastry, Girish and Askell, Amanda and Agarwal, Sandhini and Herbert-Voss, Ariel and Krueger, Gretchen and Henighan, Tom and Child, Rewon and Ramesh, Aditya and Ziegler, Daniel and Wu, Jeffrey and Winter, Clemens and Hesse, Chris and Chen, Mark and Sigler, Eric and Litwin, Mateusz and Gray, Scott and Chess, Benjamin and Clark, Jack and Berner, Christopher and McCandlish, Sam and Radford, Alec and Sutskever, Ilya and Amodei, Dario},
  editor = {Larochelle, H. and Ranzato, M. and Hadsell, R. and Balcan, M. F. and Lin, H.},
  title = {Language {Models} are {Few}-{Shot} {Learners}},
  booktitle = {Advances in {Neural} {Information} {Processing} {Systems}},
  volume = {33},
  pages = {1877--1901},
  publisher = {Curran Associates, Inc.},
  year = {2020},
}

@misc{gao_pile_2020,
  author = {Gao, Leo and Biderman, Stella and Black, Sid and Golding, Laurence and Hoppe, Travis and Foster, Charles and Phang, Jason and He, Horace and Thite, Anish and Nabeshima, Noa and Presser, Shawn and Leahy, Connor},
  title = {The {Pile}: {An} {800GB} {Dataset} of {Diverse} {Text} for {Language} {Modeling}},
  year = {2020},
  doi = {10.48550/arXiv.2101.00027},
  eprint = {2101.00027},
  archivePrefix = {arXiv},
  primaryClass = {cs.CL},
  note = {arXiv:2101.00027},
}

@article{chowdhery_palm_2023,
  author = {Chowdhery, Aakanksha and Narang, Sharan and Devlin, Jacob and Bosma, Maarten and Mishra, Gaurav and Roberts, Adam and Barham, Paul and Chung, Hyung Won and Sutton, Charles and Gehrmann, Sebastian and Schuh, Parker and Shi, Kensen and Tsvyashchenko, Sasha and Maynez, Joshua and Rao, Abhishek and Barnes, Parker and Tay, Yi and Shazeer, Noam and Prabhakaran, Vinodkumar and Reif, Emily and Du, Nan and Hutchinson, Ben and Pope, Reiner and Bradbury, James and Austin, Jacob and Isard, Michael and Gur-Ari, Guy and Yin, Pengcheng and Duke, Toju and Levskaya, Anselm and Ghemawat, Sanjay and Dev, Sunipa and Michalewski, Henryk and Garcia, Xavier and Misra, Vedant and Robinson, Kevin and Fedus, Liam and Zhou, Denny and Ippolito, Daphne and Luan, David and Lim, Hyeontaek and Zoph, Barret and Spiridonov, Alexander and Sepassi, Ryan and Dohan, David and Agrawal, Shivani and Omernick, Mark and Dai, Andrew M. and Pillai, Thanumalayan Sankaranarayana and Pellat, Marie and Lewkowycz, Aitor and Moreira, Erica and Child, Rewon and Polozov, Oleksandr and Lee, Katherine and Zhou, Zongwei and Wang, Xuezhi and Saeta, Brennan and Diaz, Mark and Firat, Orhan and Catasta, Michele and Wei, Jason and Meier-Hellstern, Kathy and Eck, Douglas and Dean, Jeff and Petrov, Slav and Fiedel, Noah},
  title = {{PaLM}: {Scaling} {Language} {Modeling} with {Pathways}},
  journal = {Journal of Machine Learning Research},
  volume = {24},
  number = {240},
  pages = {1--113},
  year = {2023},
}

@inproceedings{dao_flashattention_2022,
  author = {Dao, Tri and Fu, Dan and Ermon, Stefano and Rudra, Atri and R{\'e}, Christopher},
  editor = {Koyejo, S. and Mohamed, S. and Agarwal, A. and Belgrave, D. and Cho, K. and Oh, A.},
  title = {{FlashAttention}: {Fast} and {Memory}-{Efficient} {Exact} {Attention} with {IO}-{Awareness}},
  booktitle = {Advances in {Neural} {Information} {Processing} {Systems}},
  volume = {35},
  pages = {16344--16359},
  publisher = {Curran Associates, Inc.},
  year = {2022},
}

@inproceedings{kwon_efficient_2023,
  author = {Kwon, Woosuk and Li, Zhuohan and Zhuang, Siyuan and Sheng, Ying and Zheng, Lianmin and Yu, Cody Hao and Gonzalez, Joseph and Zhang, Hao and Stoica, Ion},
  title = {Efficient {Memory} {Management} for {Large} {Language} {Model} {Serving} with {PagedAttention}},
  booktitle = {Proceedings of the 29th {Symposium} on {Operating} {Systems} {Principles}},
  series = {{SOSP} '23},
  pages = {611--626},
  publisher = {Association for Computing Machinery},
  address = {New York, NY, USA},
  year = {2023},
  doi = {10.1145/3600006.3613165},
  isbn = {979-8-4007-0229-7},
}

@inproceedings{strubell_energy_2019,
  author = {Strubell, Emma and Ganesh, Ananya and McCallum, Andrew},
  editor = {Korhonen, Anna and Traum, David and M{\`a}rquez, Llu{\'i}s},
  title = {Energy and {Policy} {Considerations} for {Deep} {Learning} in {NLP}},
  booktitle = {Proceedings of the 57th {Annual} {Meeting} of the {Association} for {Computational} {Linguistics}},
  pages = {3645--3650},
  publisher = {Association for Computational Linguistics},
  address = {Florence, Italy},
  year = {2019},
  doi = {10.18653/v1/P19-1355},
}

@misc{patterson_carbon_2021,
  author = {Patterson, David and Gonzalez, Joseph and Le, Quoc and Liang, Chen and Munguia, Lluis-Miquel and Rothchild, Daniel and So, David and Texier, Maud and Dean, Jeff},
  title = {Carbon {Emissions} and {Large} {Neural} {Network} {Training}},
  year = {2021},
  doi = {10.48550/arXiv.2104.10350},
  eprint = {2104.10350},
  archivePrefix = {arXiv},
  primaryClass = {cs.CY},
  note = {arXiv:2104.10350},
}

@incollection{gholami_survey_2022,
  author = {Gholami, Amir and Kim, Sehoon and Dong, Zhen and Yao, Zhewei and Mahoney, Michael W. and Keutzer, Kurt},
  editor = {Thiruvathukal, George K. and Lu, Yung-Hsiang and Kim, Jaeyoun and Chen, Yiran and Chen, Bo},
  title = {A {Survey} of {Quantization} {Methods} for {Efficient} {Neural} {Network} {Inference}},
  booktitle = {Low-{Power} {Computer} {Vision}},
  pages = {291--326},
  publisher = {Chapman and Hall/CRC},
  year = {2022},
  doi = {10.1201/9781003162810-13},
}

@misc{dao_flashattention-2_2023,
  author = {Dao, Tri},
  title = {{FlashAttention}-2: {Faster} {Attention} with {Better} {Parallelism} and {Work} {Partitioning}},
  year = {2023},
  doi = {10.48550/arXiv.2307.08691},
  eprint = {2307.08691},
  archivePrefix = {arXiv},
  primaryClass = {cs.LG},
  note = {arXiv:2307.08691},
}

@inproceedings{jouppi_-datacenter_2017,
  author = {Jouppi, Norman P. and Young, Cliff and Patil, Nishant and Patterson, David and Agrawal, Gaurav and Bajwa, Raminder and Bates, Sarah and Bhatia, Suresh and Boden, Nan and Borchers, Al and Boyle, Rick and Cantin, Pierre-luc and Chao, Clifford and Clark, Chris and Coriell, Jeremy and Daley, Mike and Dau, Matt and Dean, Jeffrey and Gelb, Ben and Ghaemmaghami, Tara Vazir and Gottipati, Rajendra and Gulland, William and Hagmann, Robert and Ho, C. Richard and Hogberg, Doug and Hu, John and Hundt, Robert and Hurt, Dan and Ibarz, Julian and Jaffey, Aaron and Jaworski, Alek and Kaplan, Alexander and Khaitan, Harshit and Killebrew, Daniel and Koch, Andy and Kumar, Naveen and Lacy, Steve and Laudon, James and Law, James and Le, Diemthu and Leary, Chris and Liu, Zhuyuan and Lucke, Kyle and Lundin, Alan and MacKean, Gordon and Maggiore, Adriana and Mahony, Maire and Miller, Kieran and Nagarajan, Rahul and Narayanaswami, Ravi and Ni, Ray and Nix, Kathy and Norrie, Thomas and Omernick, Mark and Penukonda, Narayana and Phelps, Andy and Ross, Jonathan and Ross, Matt and Salek, Amir and Samadiani, Emad and Severn, Chris and Sizikov, Gregory and Snelham, Matthew and Souter, Jed and Steinberg, Dan and Swing, Andy and Tan, Mercedes and Thorson, Gregory and Tian, Bo and Toma, Horia and Tuttle, Erick and Vasudevan, Vijay and Walter, Richard and Wang, Walter and Wilcox, Eric and Yoon, Doe Hyun},
  title = {In-{Datacenter} {Performance} {Analysis} of a {Tensor} {Processing} {Unit}},
  booktitle = {Proceedings of the 44th {Annual} {International} {Symposium} on {Computer} {Architecture}},
  series = {{ISCA} '17},
  pages = {1--12},
  publisher = {Association for Computing Machinery},
  address = {New York, NY, USA},
  year = {2017},
  doi = {10.1145/3079856.3080246},
  isbn = {978-1-4503-4892-8},
}

@article{chen_eyeriss_2017,
  author = {Chen, Yu-Hsin and Krishna, Tushar and Emer, Joel S. and Sze, Vivienne},
  title = {Eyeriss: {An} {Energy}-{Efficient} {Reconfigurable} {Accelerator} for {Deep} {Convolutional} {Neural} {Networks}},
  journal = {IEEE Journal of Solid-State Circuits},
  volume = {52},
  number = {1},
  pages = {127--138},
  year = {2017},
  doi = {10.1109/JSSC.2016.2616357},
}

@article{hamerly_large-scale_2019,
  author = {Hamerly, Ryan and Bernstein, Liane and Sludds, Alexander and Solja{\v c}i{\'c}, Marin and Englund, Dirk},
  title = {Large-{Scale} {Optical} {Neural} {Networks} {Based} on {Photoelectric} {Multiplication}},
  journal = {Physical Review X},
  volume = {9},
  number = {2},
  pages = {021032},
  year = {2019},
  doi = {10.1103/PhysRevX.9.021032},
}

@article{zhou_photonic_2022,
  author = {Zhou, Hailong and Dong, Jianji and Cheng, Junwei and Dong, Wenchan and Huang, Chaoran and Shen, Yichen and Zhang, Qiming and Gu, Min and Qian, Chao and Chen, Hongsheng and Ruan, Zhichao and Zhang, Xinliang},
  title = {Photonic matrix multiplication lights up photonic accelerator and beyond},
  journal = {Light: Science \& Applications},
  volume = {11},
  number = {1},
  pages = {30},
  year = {2022},
  doi = {10.1038/s41377-022-00717-8},
}

@article{miller_device_2009,
  author = {Miller, David A. B.},
  title = {Device {Requirements} for {Optical} {Interconnects} to {Silicon} {Chips}},
  journal = {Proceedings of the IEEE},
  volume = {97},
  number = {7},
  pages = {1166--1185},
  year = {2009},
  doi = {10.1109/JPROC.2009.2014298},
}

@article{shen_deep_2017,
  author = {Shen, Yichen and Harris, Nicholas C. and Skirlo, Scott and Prabhu, Mihika and Baehr-Jones, Tom and Hochberg, Michael and Sun, Xin and Zhao, Shijie and Larochelle, Hugo and Englund, Dirk and Solja{\v c}i{\'c}, Marin},
  title = {Deep learning with coherent nanophotonic circuits},
  journal = {Nature Photonics},
  volume = {11},
  number = {7},
  pages = {441--446},
  year = {2017},
  doi = {10.1038/nphoton.2017.93},
}

@article{rizzo_massively_2023,
  author = {Rizzo, Anthony and Novick, Asher and Gopal, Vignesh and Kim, Bok Young and Ji, Xingchen and Daudlin, Stuart and Okawachi, Yoshitomo and Cheng, Qixiang and Lipson, Michal and Gaeta, Alexander L. and Bergman, Keren},
  title = {Massively scalable {Kerr} comb-driven silicon photonic link},
  journal = {Nature Photonics},
  volume = {17},
  number = {9},
  pages = {781--790},
  year = {2023},
  doi = {10.1038/s41566-023-01244-7},
}

@misc{yun_are_2020,
  author = {Yun, Chulhee and Bhojanapalli, Srinadh and Rawat, Ankit Singh and Reddi, Sashank J. and Kumar, Sanjiv},
  title = {Are {Transformers} universal approximators of sequence-to-sequence functions?},
  year = {2020},
  doi = {10.48550/arXiv.1912.10077},
  eprint = {1912.10077},
  archivePrefix = {arXiv},
  primaryClass = {cs.LG},
  note = {arXiv:1912.10077},
}

@misc{karpathy_shakespeare_2015,
  author = {Karpathy, Andrej},
  title = {char-rnn},
  year = {2015},
  howpublished = {GitHub repository},
}

@manual{nvidia_a100_datasheet_2021,
  title = {{NVIDIA A100 Tensor Core GPU Datasheet}},
  organization = {{NVIDIA Corporation}},
  year = {2021},
}

@misc{karpathy_nanogpt_2022,
  author = {Karpathy, Andrej},
  title = {{nanoGPT}: The simplest, fastest repository for training/finetuning medium-sized GPTs},
  year = {2022},
  howpublished = {GitHub repository}
}

@article{suh_soliton_2018,
  author = {Suh, Myoung-Gyun and Vahala, Kerry J.},
  title = {Soliton microcomb range measurement},
  journal = {Science},
  volume = {359},
  number = {6378},
  pages = {884--887},
  year = {2018},
  doi = {10.1126/science.aao1968}
}

@article{riemensberger_massively_2020,
  author = {Riemensberger, Johann and Lukashchuk, Anton and Karpov, Maxim and Weng, Wenle and Lucas, Erwan and Liu, Junqiu and Kippenberg, Tobias J.},
  title = {Massively parallel coherent laser ranging using a soliton microcomb},
  journal = {Nature},
  volume = {581},
  number = {7807},
  pages = {164--170},
  year = {2020},
  doi = {10.1038/s41586-020-2239-3}
}

@article{jin_combinatorial_2026,
	title = {Combinatorial optimization with {Kerr} solitons},
	volume = {12},
	doi = {10.1126/sciadv.aeb7930},
	number = {21},
	journal = {Science Advances},
	publisher = {American Association for the Advancement of Science},
	author = {Jin, Yan and Chauhan, Nitesh and Zang, Jizhao and Edwards, Brian and Chaudhari, Pratik and Aflatouni, Firooz and Papp, Scott B.},
	month = may,
	year = {2026},
	pages = {eaeb7930}
}

@article{Obrzud2017Temporal,
  author  = {Obrzud, Ewelina and Lecomte, Steve and Herr, Tobias},
  title   = {Temporal solitons in microresonators driven by optical pulses},
  journal = {Nature Photonics},
  year    = {2017},
  volume  = {11},
  number  = {9},
  pages   = {600--607},
  doi     = {10.1038/nphoton.2017.140}
}

@article{Hendry_trapping_2018,
  author = {Hendry, Ian and Chen, Wei and Wang, Yadong and Garbin, Bruno and
            Javaloyes, Julien and Oppo, Gian-Luca and Coen, St{\'e}phane and
            Murdoch, Stuart G. and Erkintalo, Miro},
  title = {Spontaneous symmetry breaking and trapping of temporal {Kerr}
           cavity solitons by pulsed or amplitude-modulated driving fields},
  journal = {Physical Review A},
  volume = {97},
  number = {5},
  pages = {053834},
  year = {2018},
  month = may,
  doi = {10.1103/PhysRevA.97.053834}
}

@article{Cole_chirped_2018,
  author = {Cole, Daniel C. and Stone, Jordan R. and Erkintalo, Miro and
            Yang, Ki Youl and Yi, Xu and Vahala, Kerry J. and Papp, Scott B.},
  title = {{Kerr}-microresonator solitons from a chirped background},
  journal = {Optica},
  volume = {5},
  number = {10},
  pages = {1304--1310},
  year = {2018},
  month = oct,
  doi = {10.1364/OPTICA.5.001304}
}

@article{Lucas_detuning_2017,
  author = {Lucas, Erwan and Guo, Hairun and Jost, John D. and Karpov, Maxim and Kippenberg, Tobias J.},
  title = {Detuning-dependent properties and dispersion-induced instabilities of temporal dissipative {Kerr} solitons in optical microresonators},
  journal = {Physical Review A},
  volume = {95},
  number = {4},
  pages = {043822},
  year = {2017},
  doi = {10.1103/PhysRevA.95.043822}
}

@article{chauhan2026tripartite,
  title = {Tripartite Kerr Soliton Ising Machine for Combinatorial Optimization},
  author = {Chauhan, Nitesh and Jin, Yan and Zang, Jizhao and Papp, Scott B.},
  journal = {Phys. Rev. Lett.},
  volume = {136},
  issue = {25},
  pages = {257301},
  numpages = {6},
  year = {2026},
  month = {Jun},
  publisher = {American Physical Society},
  doi = {10.1103/pb9b-6hyk}
}

@misc{jin2026nanophotoniccontrolcollectivemanybody,
  author = {Yan Jin and Jizhao Zang and Sarang Yeola and Alexa R. Carollo and Nitesh Chauhan and Scott B. Papp},
  title = {Nanophotonic control of collective many-body states in Kerr solitons},
  year = {2026},
  doi = {10.48550/arXiv.2604.22039},
  eprint = {2604.22039},
  archivePrefix = {arXiv},
  primaryClass = {physics.optics},
  note = {arXiv:2604.22039},
}

\end{document}